# Accelerator Science and Technology Breakthroughs, Achievements and Lessons from the Tevatron*

V. Shiltsev, FNAL, Batavia, IL 60510, USA[#]

*To Fermilab staff who made the Tevatron collider a great success.*

*Abstract*

For almost a quarter of a century, the Tevatron proton-antiproton collider was the centerpiece of the world's high energy physics program – beginning operation in December of 1985 until it was overtaken by LHC in 2011. The aim of this unique scientific instrument was to explore the elementary particle physics reactions with center of mass collision energies of up to 1.96 TeV. The initial design luminosity of the Tevatron was $10^{30} cm^{-2} s^{-1}$, however as a result of two decades of upgrades, the accelerator has been able to deliver 430 times higher luminosities to each of two high luminosity experiments, CDF and D0. The Tevatron will be shut off September 30, 2011. The collider was arguably one of the most complex research instruments ever to reach the operation stage and is widely recognized for many technological breakthroughs and numerous physics discoveries. In this John Adams lecture, I briefly present the history of the Tevatron, major advances in accelerator physics, and technology implemented during the long quest for better and better performance. Lessons learned from our experience are also discussed.

## *Introduction: History and Performance*

As a representative of Fermilab, I have particular pleasure using this opportunity to present accelerator physics and technology achievements of the Tevatron collider as part CERN's John Adams lecture series. One reason is that CERN's Large Hadron Collider has taken the energy frontier leadership from the Tevatron – first, in the colliding beam energy (on November 29, 2009) and recently in terms of colliding beam luminosity (on April 22, 2011.) This makes it a natural time to look back and reflect on how the Tevatron successes paved the way to the current frontier machine. Another reason is the tradition of very tight and friendly "cooperation-competition" between CERN and Fermilab. One can trace this all the way back to the time of our great "founding fathers" such as John Adams and Bob Wilson. Those two, together with Soviet scientist Gersh Budker of Novosibirsk Institute of Nuclear Physics and "Pief" Panofsky of SLAC shaped the world of large scale, energy frontier particle accelerators as we know it now. In that regard, they were direct descendants of the inventor of the cyclotron, Nobel Prize winner Ernest Lawrence who "started that all." I and most of today's accelerator community are therefore, "grandchildren" or "great-grandchildren" of the "magnificent four." (Budker was the PhD thesis advisor of my PhD thesis advisor). Many historical reminiscences can be found in the 2010 John Adams seminar lecture by E.Wilson [1] and recent book on the Tevatron history [2].

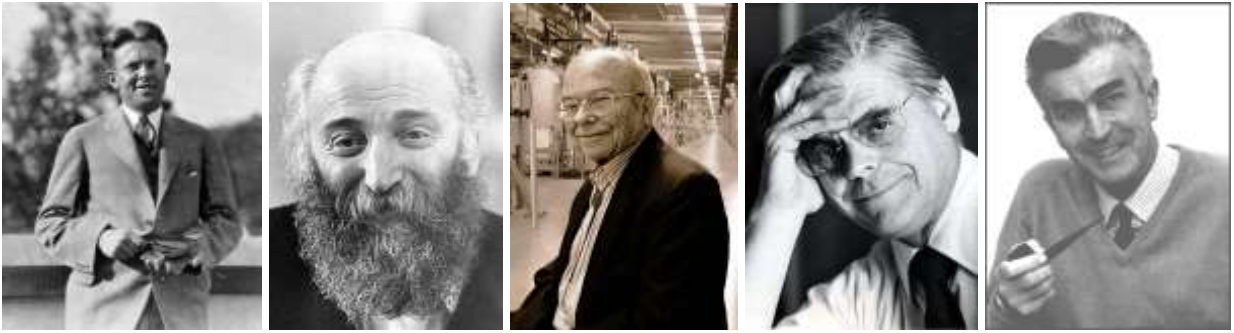

Figure 1: People that shaped the world of particle accelerators (left to right): Ernest Lawrence (1901-1958), Gersh Budker (1918-1977), Wolfgang Panofsky (1919-2007), Robert R Wilson (1914-2000), John Adams (1920-1984).

The Tevatron was conceived by Bob Wilson [3] to double the energy of the Fermilab complex from 500 GeV to 1000 GeV. The original name, the "Energy Saver/Doubler", reflected this mission and the accrued benefit of reduced power utilization through the use of superconducting magnets. The introduction of superconducting magnets in a large scale application allowed the (now named) Tevatron to be constructed with the same circumference of 6.3 km, and to be installed in the same tunnel as the original Main Ring proton synchrotron which would serve as its injector (at 150 GeV). Superconducting magnet development was initiated in the early 1970's and ultimately produced successful magnets, leading to commissioning of the Tevatron in July 1983.

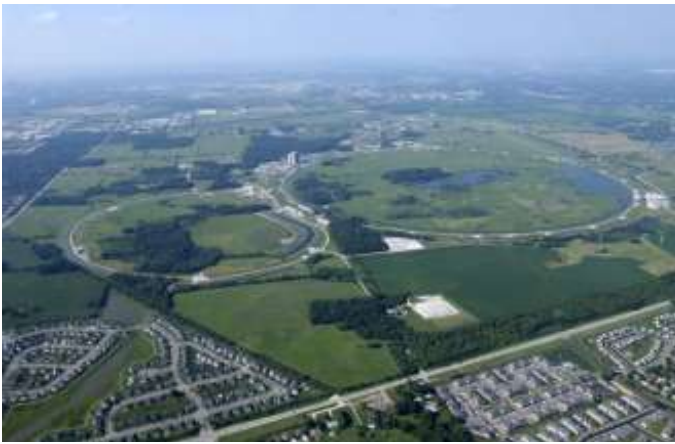
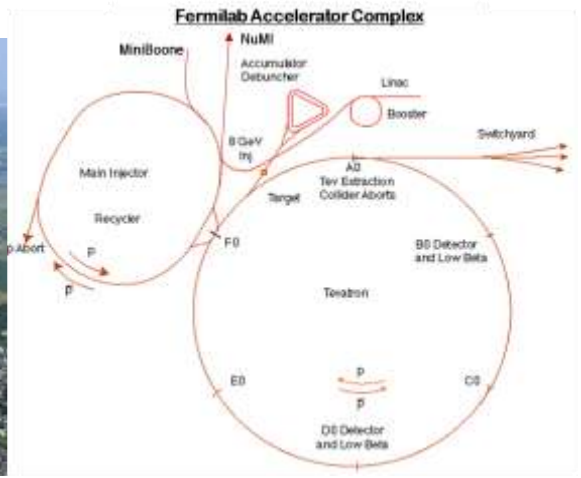

Figure 2: Aerial view of Fermilab (left) and layout of its accelerator complex (right).

In 1976 D.Cline et al. proposed a proton-antiproton collider with luminosities of about $10^{29}$ cm$^{-2}$sec$^{-1}$ at Fermilab [4] or at CERN, based on the conversion of an existing accelerator into a storage ring and construction of a new facility for the accumulation and cooling of approximately $10^{11}$ antiprotons per day. The motivation was to discover the intermediate vector bosons. The first antiproton accumulation facility was constructed at CERN and supported collisions at 630 GeV(center-of-mass) in the modified SPS synchrotron, where the W and Z particles were discovered in 1983. Meanwhile, in 1978 Fermilab decided that proton-antiproton collisions would be supported in the Tevatron, at a center-of-mass energy of 1800 GeV and that an Antiproton Source facility would be constructed to supply the flux of antiprotons needed for a design luminosity of $1\times10^{30}$ cm$^{-2}$sec$^{-1}$.

The Tevatron as a fixed target accelerator was completed in 1983 [5]. The Antiproton Source [6] was completed in 1985 and first collisions were observed in the Tevatron using some operational elements of the CDF detector (then under construction) in October 1985. Initial operation of the Collider for data taking took place during a period from February through May of 1987. A more extensive run took place between June 1988 and June 1989, representing the first sustained operation at the design luminosity. In this period of operation a total of 5 pb$^{-1}$ were delivered to CDF at 1800 GeV (center-of-mass) and the first western hemisphere W's and Z's were observed. The initial operational goal of $1\times10^{30}$ cm$^{-2}$sec$^{-1}$ luminosity was exceeded during this run. Table I summarizes the actual performance achieved in the 1988-89 run. (Short runs at √s = 630 GeV and √s = 1020 GeV also occurred in 1989.)

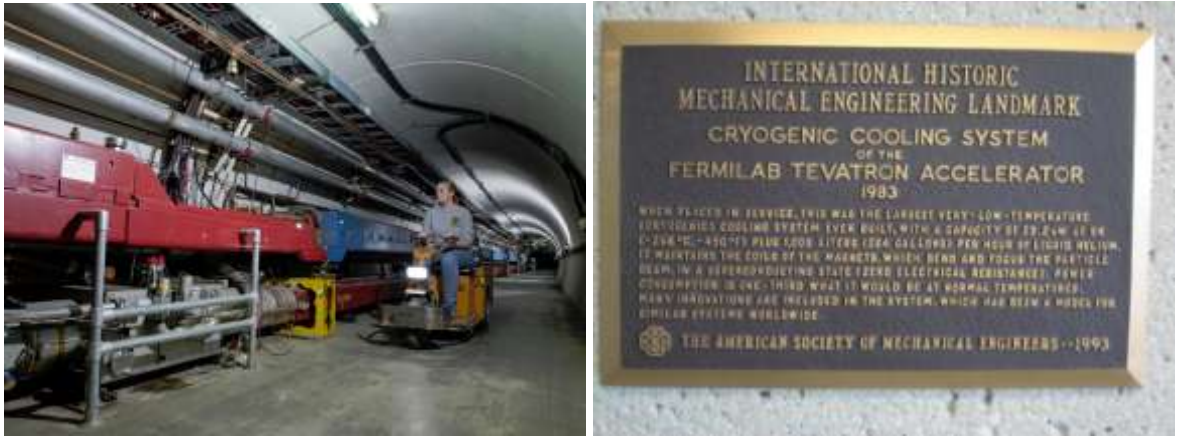

Figure 3: (left) In the the Tevatron/Main Ring tunnel in 2000's – larger magnets are remnants of the old 500 GeV Main Ring, the Tevatron accelerator magnets are the smaller ones, placed almost on the floor (10.5 inches between the center of the beampipe and the floor level); (right) the Tevatron cryogenic plant was recognized as an International Mechanical Engineering landmark as when it was built in early 1980's it was the largest cryo-factory in the world, delivering about 23kW of cooling power at the liquid helium temperature of 5 Kelvin.

In the early to mid-1990's a number of improvements were implemented to prepare for operation of Collider Run I (August of 1992 through February 1996):

Electrostatic separators aimed at mitigating the beam-beam limitations by placing protons and antiprotons on separate helical orbits, thus allowing an increase in the number of bunches and proton intensity: twenty-two, 3 m long, electrostatic separators operating at up to ±300 kV across a 5 cm gap were installed in the Tevatron by 1992. During Run II (2001-2011), 4 additional separators were installed to improve separation at the nearest parasitic crossings.

Low beta systems which ultimately allowed operations with $\beta^*$ less than 30 cm: The 1988-89 Run did not have a matched insertion for the interaction region at B0 (where CDF was situated). Two sets of high performance quadrupoles were developed and installed at B0 and D0 (which came online for Run I in 1992).

Cryogenic cold compressors lowered the operating Helium temperature by about 0.5 K, thereby allowing the beam energy to be increased to 1000 GeV, in theory. In operational practice 980 GeV was achieved.

Antiproton Source improvements: A number of improvements were made to the stochastic cooling systems in the Antiproton Source in order to accommodate higher antiproton flux generated by continuously increasing the proton intensity on the antiproton production target. Improvements included the introduction of transverse stochastic cooling into the Debuncher and upgrades to the bandwidth of the core cooling system. These improvements supported an accumulation rate of $7\times10^{10}$ antiprotons per hour.

Run I consisted of two distinct phases, Run Ia which ended in May of 1993, and Run Ib which was initiated in December of 1993. The 400 MeV linac upgrade (from the initial 200 MeV) was implemented between Run Ia and Run Ib with the goal of reducing space-charge effects at injection energy in the Booster and provide higher beam brightness at 8 GeV. As a result, the total intensity delivered from the Booster increased from roughly $3\times10^{12}$ per pulse to about $5\times10^{12}$. This resulted in more protons being transmitted to the antiproton production target and, ultimately, more protons and antiprotons in collision in the Tevatron.

Run I ultimately delivered a total integrated luminosity of 180 pb$^{-1}$ to both CDF and D0 experiments at $\sqrt{s}$ = 1800 GeV. By the end of the run the typical luminosity at the beginning of a store was about $1.6\times10^{31}$ cm$^{-2}$sec$^{-1}$, a 60% increase over the Run I goal. (A brief colliding run at $\sqrt{s}$ = 630 GeV also occurred in Run I.)

In preparation for the next Collider run, construction of the Main Injector synchrotron and Recycler storage ring was initiated and completed in the spring of 1999 with the Main Injector initially utilized in the last Tevatron fixed target run.

The Main Injector was designed to significantly improve antiproton performance by replacing the Main Ring with a larger aperture, faster cycling machine [7]. The goal was a factor of three increase in the antiproton accumulation rate (to $2\times10^{11}$ per hour), accompanied by the ability to obtain 80% transmission from the Antiproton Source to the Tevatron from antiproton intensities up to $2\times10^{12}$. An antiproton accumulation rate of $2.5\times10^{11}$ per hour was achieved in Collider Run II, and transmission efficiencies beyond 80% for high antiproton intensities were routine.

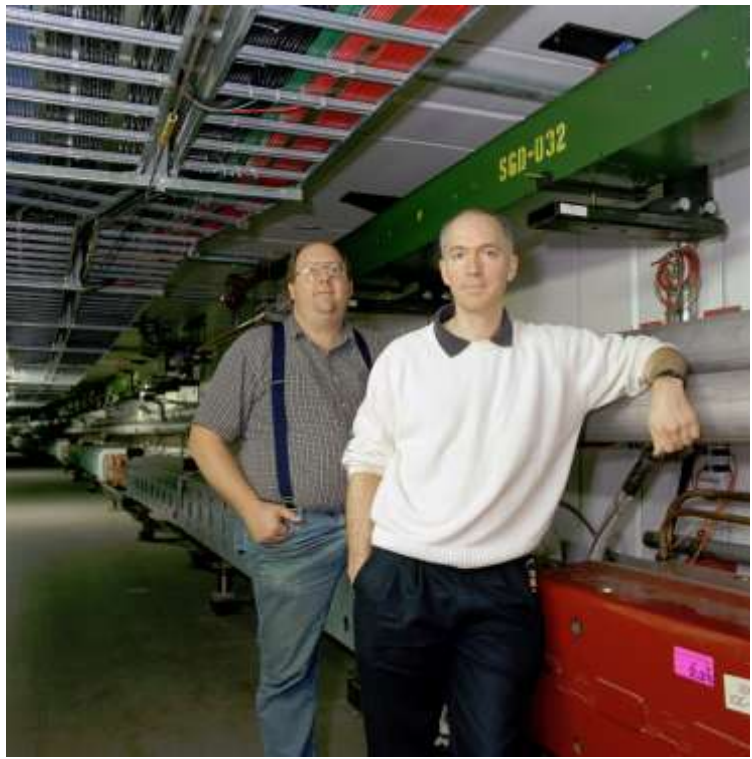

Figure 4: Designers of the 8 GeV permanent magnet Recycler Ring, Drs. Gerald Jackson(left) and William Foster, in the Main Injector tunnel. The Recycler is seen as a smaller size ring under the ceiling, while normal conducting Main Injector fast cycling synchrotron magnets are near the floor.

The Recycler was added to the Main Injector Project midway through the project (utilizing funds generated from an anticipated cost under run.) As conceived, the Recycler would provide storage for very large numbers of antiprotons (up to $6\times10^{12}$) and would increase the effective production rate by recapturing unused antiprotons at the end of collider stores [8]. The Recycler was designed with stochastic cooling systems but R&D in electron cooling was initiated in anticipation of providing improved performance. Antiproton intensities above $5\times10^{12}$ were ultimately achieved although routine operation was eventually optimized around $4\times10^{12}$ antiprotons. Recycling of antiprotons was never implemented, as the most efficient use of the new machine was operationally found to be as an additional storage ring to accumulate and cool antiprotons from the Antiproton Source and optimal reformatting the beam for injection to Tevatron.

Table 1: Achieved performance parameters for Collider Runs I and II (typical values at the beginning of a store.)

|  | 1988-89 Run | Run Ib | Run II |  |
| --- | --- | --- | --- | --- |
| Energy (center of mass) | 1800 | 1800 | 1960 | GeV |
| Protons per bunch | 7.0 | 23 | 29 | $\times 10^{10}$ |
| Antiprotons per bunch | 2.9 | 5.5 | 8.1 | $\times 10^{10}$ |
| Bunches in each beam | 6 | 6 | 36 |  |
| Total antiprotons | 17 | 33 | 290 | $\times 10^{10}$ |
| Proton emittance (rms, norm.) | 4.2 | 3.8 | 3.0 | $\pi$ μm |
| Antiproton emittance (rms, norm.) | 3 | 2.1 | 1.5 | $\pi$ μm |
| $\beta^*$ at the IPs | 55 | 35 | 28 | cm |
| Luminosity (typical, start of store) | 1.6 | 16 | 350 | $10^{30}$cm$^{-2}$s$^{-1}$ |
| Luminosity integral | $5\cdot10^{-3}$ | 0.18 | 11.9 | fb$^{-1}$ |

The Main Injector (MI) and Recycler (RR) completed the Fermilab accelerator complex development - see the ultimate scheme of operational accelerators in Fig.2 - and constituted the improvements associated with Collider Run II [8]. The luminosity goal of Run II was $8\times10^{31}$ cm$^{-2}$sec$^{-1}$, a factor of five beyond Run I. However, incorporation of the RR into the Main Injector Project was projected to provide up to an additional factor of 2.5.

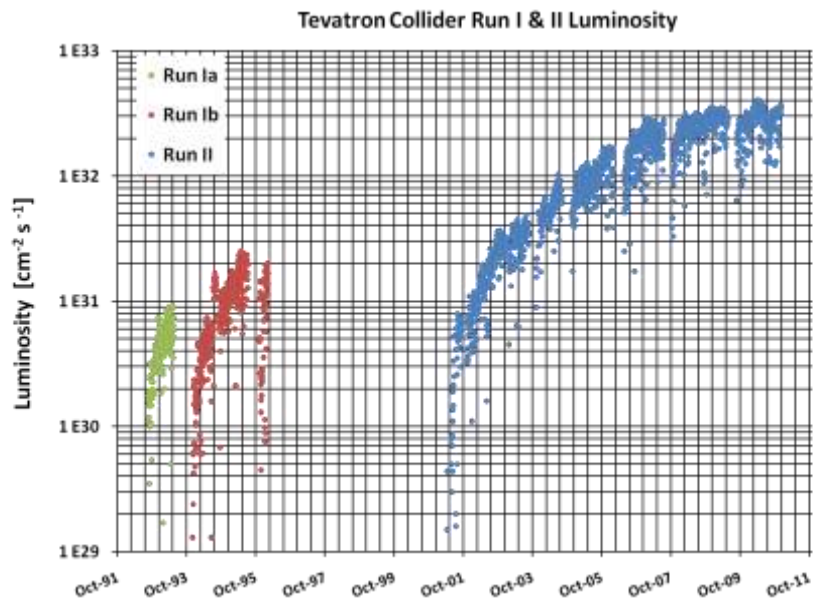

Figure 5: Initial luminosity for all Collider stores

Run II was initiated in March of 2001 and continued through September 2011. A number of difficulties were experienced in the initial years of operation. These were ultimately overcome through experience accumulated in the course of operation and the organization and execution of a "Run II Upgrade Plan". At the end of the Run II, typical Tevatron luminosities were well in excess of $3.4\times10^{32}$ cm$^{-2}$ sec$^{-1}$, with record stores exceeding $4.3\times10^{32}$ cm$^{-2}$sec$^{-1}$ – see the achieved performance parameters in Table I.

The Collider performance history (see Fig.5) shows the luminosity increases occurred after numerous improvements, some were implemented during operation and others were introduced during regular shutdown periods. They took place in all accelerators and addressed all parameters affecting luminosity – proton and antiproton intensities, emittances, optics functions, bunch length, losses, reliability and availability, etc. Analysis [10] indicates that as the result of some 32 major improvements in 2001-2011, the peak luminosity has grown by a factor of about 54 from $L_i \approx 8\times10^{30}$ cm$^{-2}$s$^{-1}$ to $L_f \approx 430\times10^{30}$ cm$^{-2}$s$^{-1}$, or about 13% per step on average. The pace of the Tevatron luminosity progress was one of the fastest among high energy colliders [11].

In general, the complex percentages, i.e. "*N*% gain per step, step after step, with regular periodicity" explain the *exponential* growth of the luminosity

$$L(t_0+T)=L(t_0)\times e^{T/C} \quad (1).$$

The pace of the luminosity progress was not always constant. As one can see in Fig.5, the Collider Run II luminosity progress was quite fast with $C\approx0.7$ year in the startup period from 2001 to mid-2002; stayed on a steady exponential increase path with $C\approx2.0$ years from 2002 until 2007, and significantly slowed down afterward, during the "stabilized operation" period, with $C\approx8.6$ years. Other high energy particle colliders show very similar features of the luminosity evolution [11]: usually, the very fast progress during the start-up period is followed by extended period of time with exponential growth of the performance which fades when the all the possibilities and ideas for further improvements are fully explored and luminosity stabilizes at its ultimate level. The coefficients *C* for various colliding facilities during most active periods of operation vary from 1.6 to 4.4.

The evolution of the performance of continuously improving facilities where every next step brings *x-fold* improvement on top of previous improvement can be further simplified in an approximate formulae, also called "*CPT theorem for accelerators*" [11]:

$$C \cdot P = T \quad (2)$$

where the factor $P=ln$(luminosity) is the "*performance*" gain over time interval *T*, and *C* is a machine dependent coefficient equal to average time needed to increase the performance (in the case of colliders – luminosity) by $e=2.71\ldots$ times, or boost the "performance" *P* by 1 unit. Both, *T* and *C* have dimension of time, and the coefficient *C* can called and has the meaning of the "*empirical complexity*" of the machine, as it directly indicates how hard or how easy was/is it to push the performance of the individual machine. In general, one can rightfully guess the complexity *C* should be dependent on how well understood are the physics and technology of the machine, type of particles, efforts and resources invested into operation and upgrades of the system, number of elements and subsystems. Such a behavior was found to be natural for many complex scientific and technological systems, and the *complexity* coefficients of some of them are listed in Table II.

TABLE II: Progress rates ("complexities") of scientific and technical systems [11].

|  | *C, years* | *Interval* | *Comment* |
|---|---|---|---|
| Fastest Computers | *1.6 ±0.1* | 1993-2010 | http://www.top500.org/ |
| Luminosity of Tevatron | *2.0 ±0.2* | 2002-2007 | see Figure 5 |
| Fusion Reactors | *2.4 ±0.2* | 1969-1999 | *Fusion Triple Product* |
| Transistors per IC | *2.7 ±0.05* | 1971-2009 | *Moore's Law* |
| Galaxies Surveyed | *3.0 ±0.1* | 1985-1990 | See refs in [11] |
| Light per LED | *3.3 ±0.1* | 1969-2000 | *Heitz's Law* |
| Most powerful lasers | *3.3 ±0.5* | 1975-2000 | http://laserstars.org/ |
| Protein Structures | *4.2 ±0.2* | 1976-2010 | http://www.pdb.org/ |
| Exoplanets Search | *4.2 ±0.3* | 1991-2010 | NASA data |
| Energy of accelerators | *5.2 ±0.3* | 1930-1990 | *Livingston plot* |
| Protons accelerated | *7.2 ±0.6* | 1960-2009 |  |

Further details of the accelerator complex evolution, basic operation of each Fermilab's machine and luminosity performance can be found in [10] and references therein. A detailed account of the first decades of the Fermilab's history can be found in the book [2].

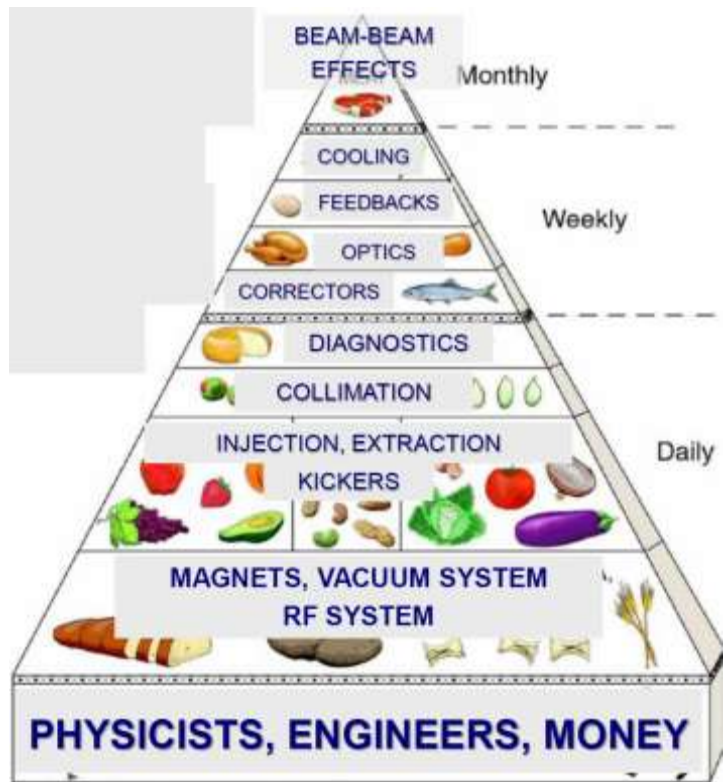

Figure 6: Pyramid of a "healthy accelerator": from strong team and reliable infrastructure in the base to the most complicated issues of beam physics at the top.

## *Major accelerator physics and technology achievements*

The presentation below will not fully cover all the important things required for such a complex system as the Tevatron Collider to operate at the peak of its performance. As all of us appreciate, "healthy" accelerators – cartoonishly depicted in Fig.6 - require very strong teams of engineers, technicians, support services, managers and administrators, contractors; it requires financial support and political backup by society; it also requires immense attention to details, sometimes "babysitting" the machine, creativity in resolving numerous day-by-day issues, courage to make important decisions to overcome difficulties, etc. All that goes beyond the topic of this talk, and the author – being just an ordinary accelerator physicist – leaves the important task of covering all that to someone else who is more knowledgeable in these issues.

So, below we give just a few examples of numerous achievements in the field of accelerator technology and beam physics which were initiated and implemented during more than three decades of Tevatron history. For more details, readers can refer to the list of references at the end of this article, to articles published in the *JINST Special Issue* [12], or to the book [13].

*Tevatron Superconducting Magnets:*
Superconducting magnets define the Tevatron, the first synchrotron built with this technology [14]. The Tevatron SC magnets experience paved the way for other colliders: HERA, RHIC, LHC – see Figs.7 and 8. Issues that had to be addressed included conductor strand and cable fabrication, coil geometry and fabrication, mechanical constraint and support of the coils, cooling and insulation, and protection during quenches. The coil placement, and hence magnetic field uniformity at the relative level of few $10^{-4}$, had the biggest effect on the accelerator performance. The magnets, designed in the 70's, performed beautifully over the years, though offered us a number of puzzles to resolve for optimal operation, like "chromaticity snap-back" effect [15] and coupling due to the cold-mass sagging [16].

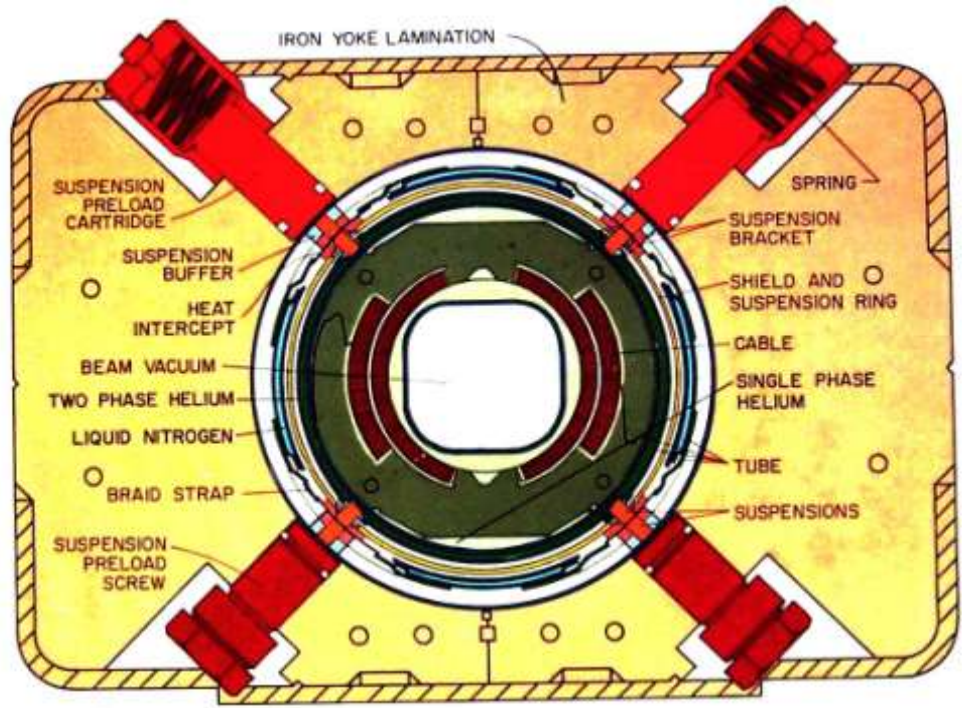

Figure 7: Cross-section of the Tevatron superconducting dipole magnet.

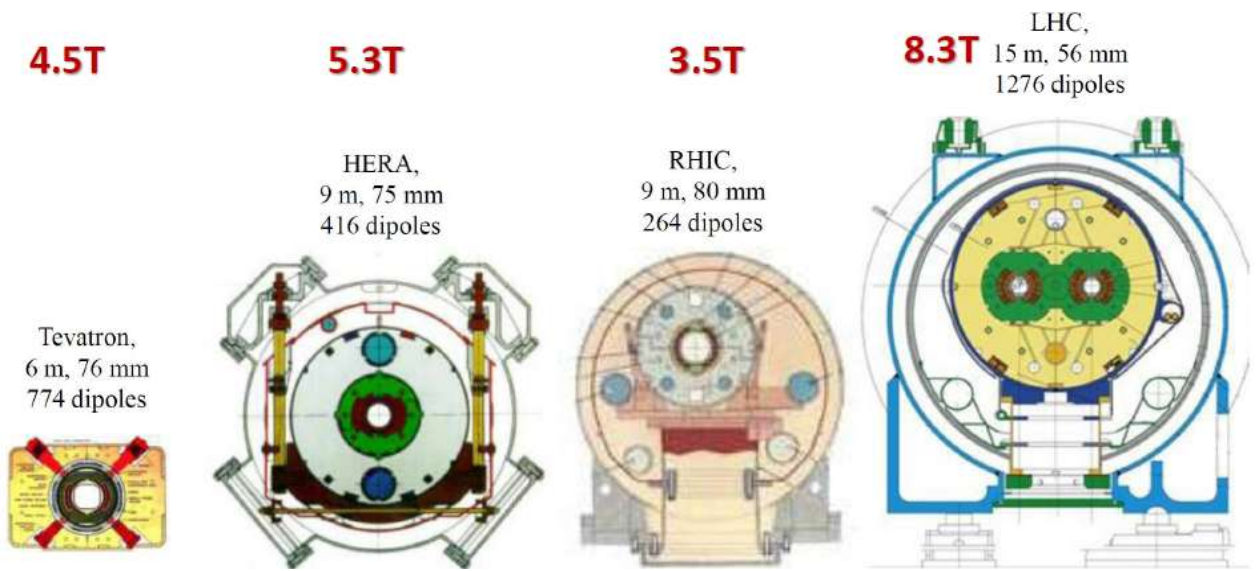

Figure 8: Superconducting dipole magnets for high energy hadron colliders: Tevatron (NbTi, warm-iron, small He plant, 4.5K), HERA (NbTi, Al collar, cold iron), RHIC (simple and economical design) and LHC (2K super fluid He, double bore) - courtesy of Dr. Alexander Zlobin, (Fermilab).

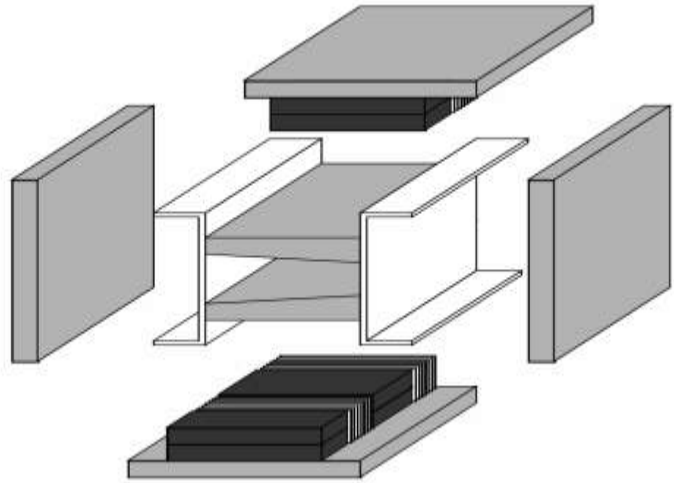

Figure 9: Recycler permanent magnet gradient dipole components shown in an exploded view. For every 4" wide brick there is a 0.5" interval of temperature compensator material composed of 10 strips.

*Recycler Permanent Magnets:*

The Recycler was the first high energy accelerator ever built with permanent magnets. It also was arguably the cheapest accelerator built (per GeV) and it employed 362 gradient dipole and 109 quadrupole magnets made of SrFe (peak field of about 1.4T) [17]. The biggest challenge was to compensate for the intrinsic temperature coefficient of the ferrite field of -0.2% per $^o$C. This was canceled down to the required 0.01%/°C by interspersing a thin NiFe "compensator alloy" strip between the ferrite bricks the pole tips. The magnetic field drifted (logarithmically slow) by a minuscule 0.04% over many years of operation [18].

*Production and Stochastic Cooling of Antiprotons :*

Stochastic cooling system technology at Fermilab expanded considerably on the initial systems developed at CERN. A total of 25 independent cooling systems were utilized for increasing phase space density of 8 GeV antiprotons in three Fermilab antiproton synchrotrons: Accumulator, Debuncher, and Recycler. The development of the systems at Fermilab have been ongoing since the early days of commissioning in 1985, and greatly benefited from improvements of liquid He-cooled pickup and kickers, preamplifiers, power amplifiers and recursive notch filters, better signal transmission and equalizers [19]. Together with increased proton intensity on the target, better targetry, and electron cooling in the Recycler, that led to stacking rates of antiprotons in excess of $28 \times 10^{10}$ per hour (q world record – see Fig.10); stacks in excess of $300 \times 10^{10}$ were accumulated in the Accumulator ring and $600 \times 10^{10}$ in the Recycler [20]. Note, that over the years, some $10^{16}$ antiprotons were produced and accumulated at Fermilab, that is about 17 nanograms and more than 90% of the world's total man-made production of nuclear antimatter. Very useful by-products of those developments were the bunched beam stochastic cooling system, implemented at RHIC [21], and multi-GHz Schottky monitors, successfully employed for multi-bunch non-invasive diagnostics of (simultaneously many) beam parameters in the Tevatron, Recycler and the LHC [22] (see Fig.11).

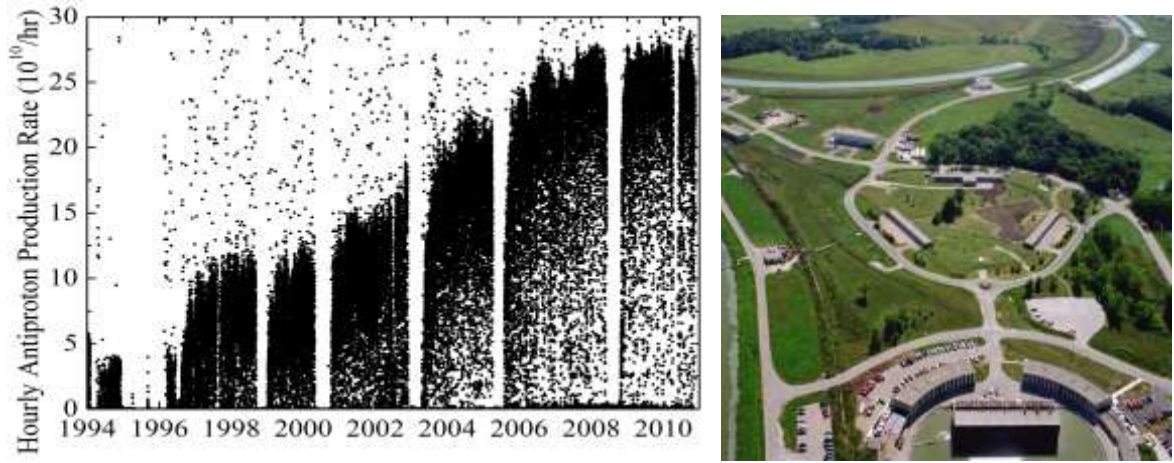

Figure 10: (left) Average hourly antiproton accumulation rate achieved during the Tevatron Collider Runs since 1994, including production in the Antiproton Source and storage in RR; (right) aerial photo of the Fermilab's Antiproton Sources accelerators: triangular shape Accumulator and Debuncher rings and (part of) the Recycler Ring, faint straight lines on the surface connecting some locations of the rings indicate links between stochastic cooling pickups and kickers.

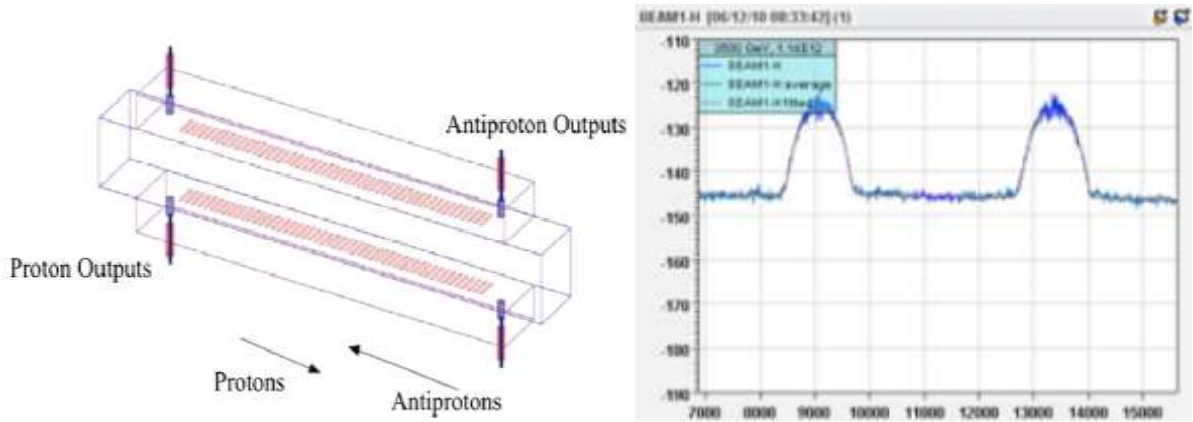

Figure 11: (left) Slotted waveguide structure of the 1.7 GHz Schottky monitors employed in the Tevatron; (right) LHC beam Schottky spectrum at frequencies around 4.8GHz (from Ref. [22]).

*High Energy Electron Cooling of Antiprotons:*

One of the most critical elements in the evolution of Run II was the successful introduction of high energy electron cooling [23] into the Recycler during the summer of 2005. Prior to the electron cooling luminosities had approached, but not exceeded, $1 \times 10^{32}$ cm$^{-2}$sec$^{-1}$, while the cooling opened the possibility for several times higher, record performance.

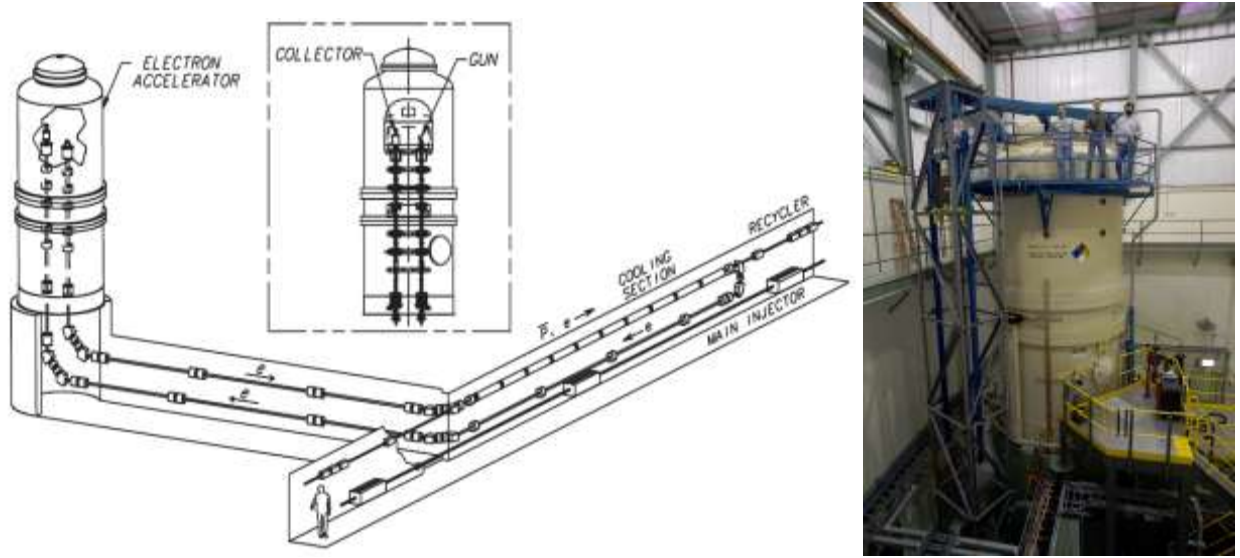

Figure 12: Schematic layout of the Recycler electron cooling system (left) and photo of the 4.5MeV Pelletron HV electrostatic accelerator – the key element of the electron cooling system.

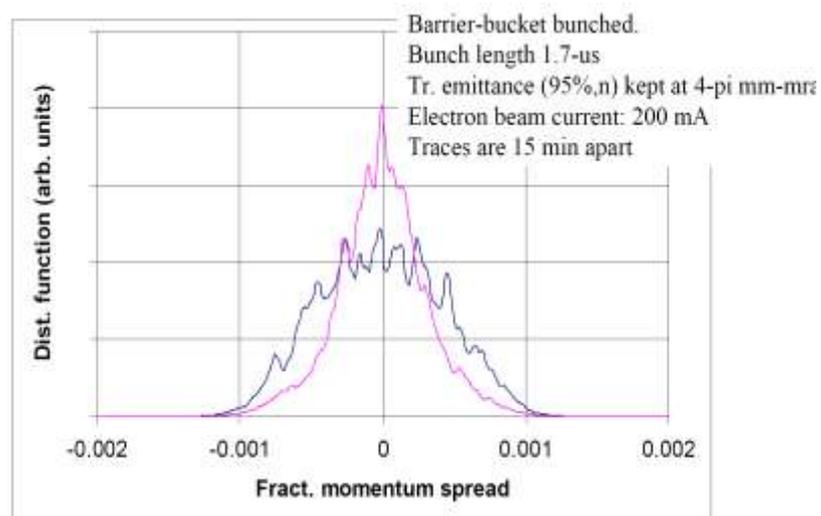

Figure 13: Schottky signal spectrum of the antiproton beam in the Recycler indicating almost two-fold reduction of the beam momentum spread after just 15 minutes of electron cooling (courtesy Dr. Sergei Nagaitsev, Fermilab).

The project overcame not only the great challenge of operating a 4.4 MeV Pelletron accelerator in the recirculation mode with up to 1A electron beam, but also resolved the hard issue of high quality magnetized electron beam transport through a non-continuous magnetic focusing beamline [24].

*Slip-Stacking and Barrier-Bucket RF Manipulations:*
Two innovative methods of longitudinal beam manipulation were developed and implemented in operation and were crucial for the success of the Tevatron Run II: a) multi-batch slip stacking [25] that allowed to approximately double the 120 GeV proton bunch intensity for antiproton production; b) the RF barrier-bucket system with rectangular 2kV RF voltage pulses [26] allowed for a whole new range of antiproton beam manipulation in the Recycler including operational "momentum mining" of antiprotons for the Tevatron shots [27] – see Figs.14 and 15.

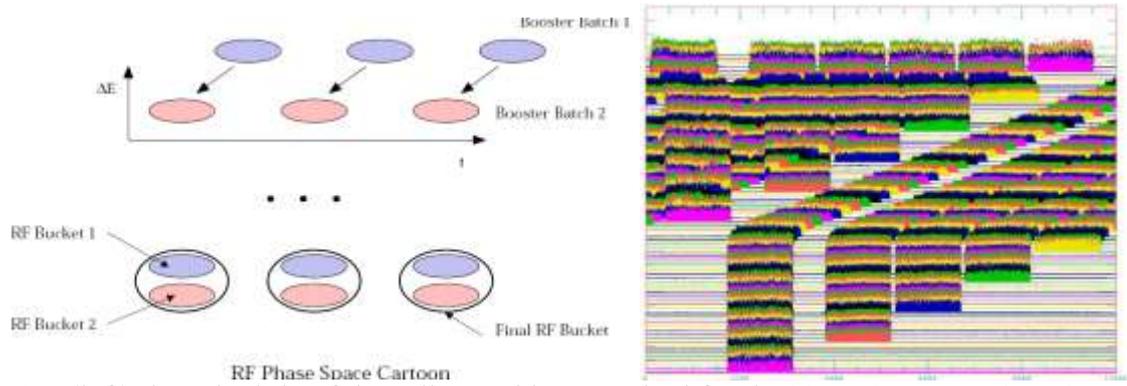

Figure 14: (left) the principle of the "slip-stacking" method for doubling the proton bunch intensity; (right) mountain range plot showing 11 batch slip stacking process in the FNAL Main Injector, horizontal scale is 10 μsec.

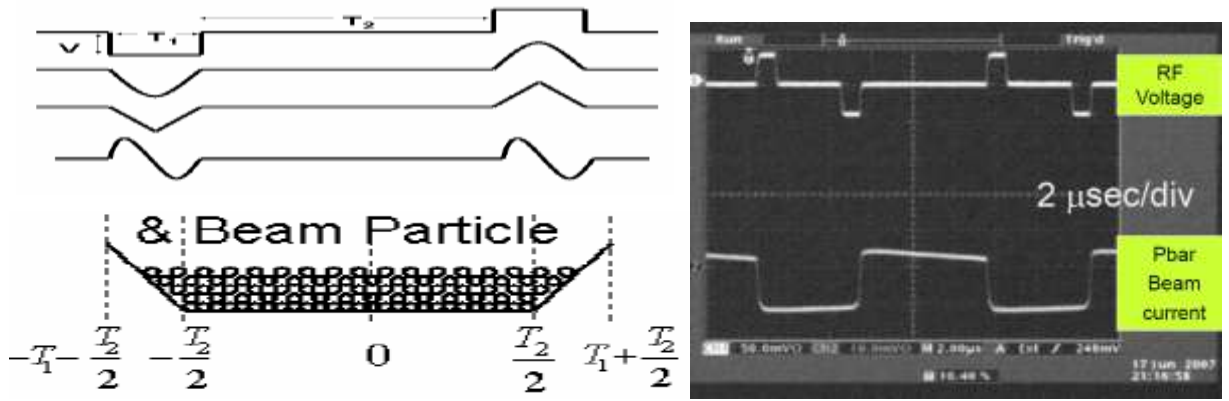

Figure 15: (left) the principle of the RF "barrier bucket" method for beam manipulation; (right) scope traces of the +- 2kV RF barrier buckets in the Recycler and antiproton beam current profile at one of the stages of operation (prior to so-called "momentum mining").

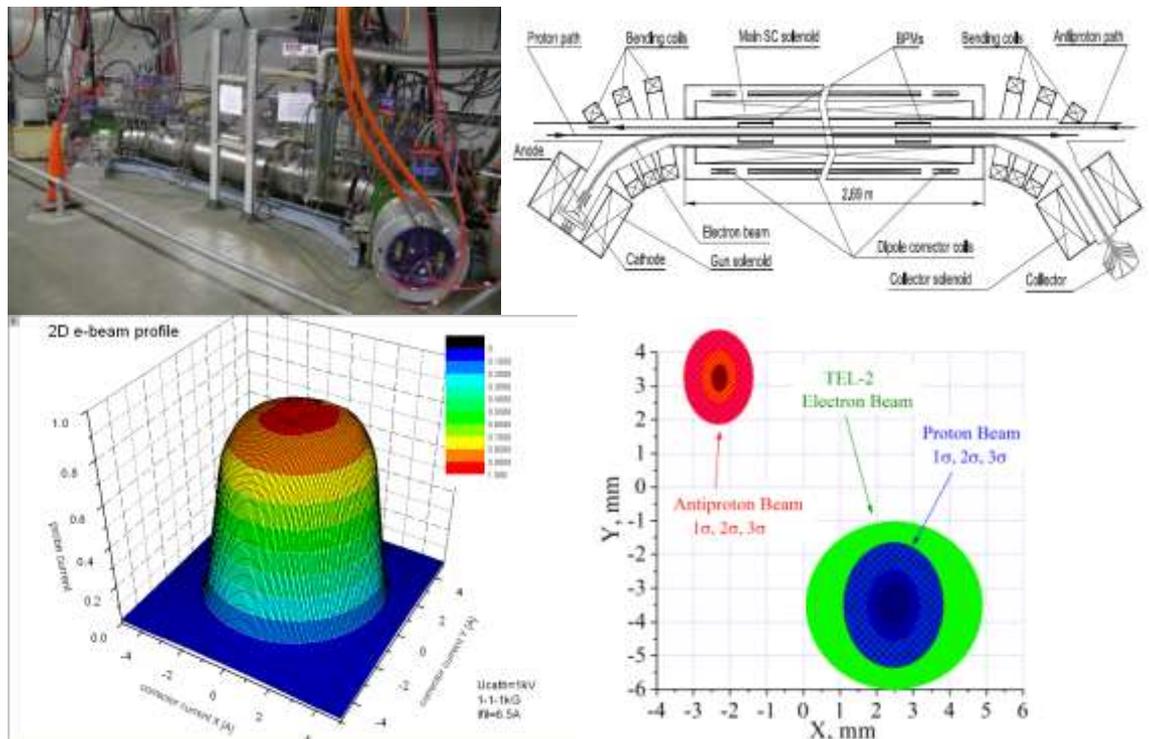

Figure 16: (top left) Photo of Tevatron Electron Lens #2 – TEL2, installed at A11 location of the Tevatron; (top right) General layout of the Tevatron Electron Lens; (bottom left) electron current profile for the TEL#2 while operating in the regime of beam-beam compensation; (bottom right)

relative position and sizes of electron, proton and antiproton beams in the configuration for compensation of the beam-beam effects (Refs. [29,30]).

*Electron Lenses for Beam-Beam Compensation:*

Electron lenses [28,29] are a novel accelerator technology used for compensation of the long-range beam-beam effects in the Tevatron [30,31], operational DC beam removal out of the Tevatron abort gaps [32], and, recently, for hollow electron beam collimation demonstration [33]. Two electron lenses were built and installed in A11 and F48 locations of the Tevatron ring. They use an 1-3 A, 6-10 kV e-beam generated at the 10-15 mm diameter thermionic cathodes immersed in a 0.3T longitudinal magnetic field and aligned onto the (anti)proton beam orbit over about 2 m length inside a 6T SC solenoid.

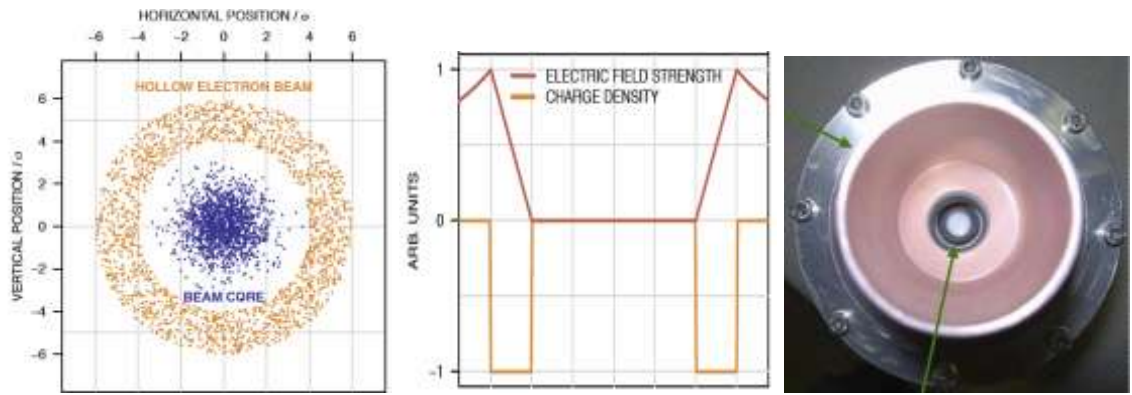

Figure 17: (left) The method of the Hollow Electron Beam Collimation (HEBC) – low energy magnetized electron beam current surrounds the high energy beam of (anti)protons in the Tevatron; (center) there are no electric or magnetic fields inside the electron beam hole, and only halo (anti)protons are affected by electrons and diffuse quickly to collimators; (right) photo of the hollow electron gun cathode, installed in TEL2 and employed for successful demonstration of HEBC in the Tevatron [33].

We should emphasize that the Tevatron Collider Runs not only delivered excellent performance (integrated luminosity), but also greatly advanced the whole accelerator field by studies of beam-beam effects [34], crystal collimation [35], electron cloud [36] and beam emittance growth mechanisms [37], new theories of beam optics [38], intra-beam scattering [39] and instabilities [40], sophisticated beam-beam and luminosity modeling [41] and more efficient beam instrumentation [42].

## Lessons from Tevatron

The Tevatron collider program will end on September 30, 2011. The machine has worked extremely well for 25 years. It has enabled CDF and D0 to discover the top quark and observe important features of the standard model for the first time. The Collider has greatly advanced accelerator technology and beam physics. Its success is a great tribute to the Fermilab staff.

We can draw several lessons from Tevatron's story:

a) one can see that exchange of ideas and methods and technology transfer helps our field: Fermilab scientists learned and borrowed a great deal of knowledge from ISR and SppS accelerators, and in turn, Tevatron's technology, techniques and experience have been successfully applied to HERA, RHIC and LHC;

b) operation of such complex systems as hadron colliders require us to be persistent and stubborn, pay close attention to details, and not count on "silver bullets" but instead be ready to go through incremental improvements.

c) it has taught us to be flexible, look for all possibilities to increase luminosity and to not be afraid to change plans if experience shows the prospects diminishing due to the complexity of machines and often unpredictability of the performance limits. Expectations management is very important.

d) operational difficulties not only generate strain, but also inspire and exalt creativity in the entire team of scientists and engineers, managers and technicians, support staff and collaborators.

Hence what I and many of us can say about the Tevatron years "it was the best time of my life".

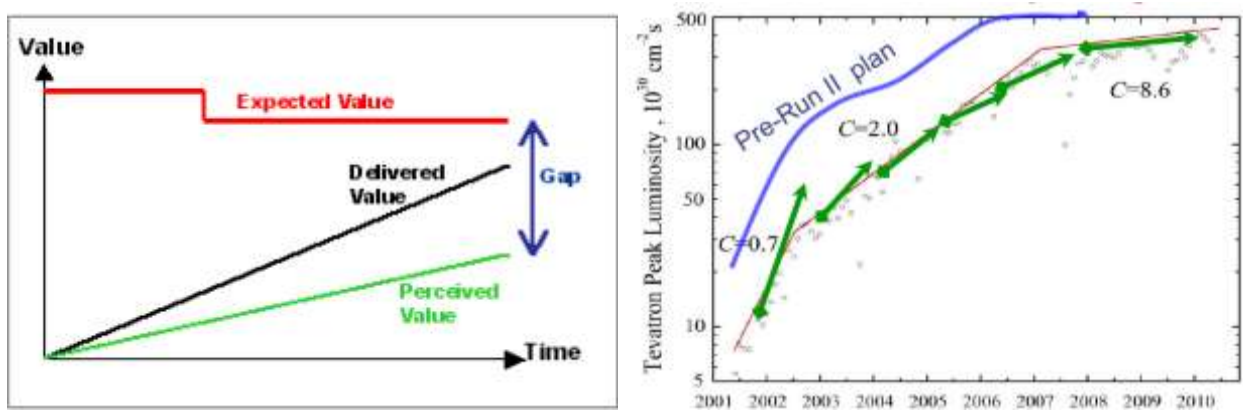

Figure 19: (left) explanatory cartoon on importance of the expectations management; (right) the Tevatron Collider plans changed as soon as we learned from difficulties at the beginning of Run II, understood the reasons for the initial slow start, figured out the methods to resolve the issues and implemented the "Run II Luminosity Upgrade" plan to achieve superb performance.

## *Acknowledgements*